\documentclass[prl,aps,twocolumn,%
showpacs,preprintnumbers,amsmath,amssymb]{revtex4-1}
\usepackage[utf8]{inputenc}
\usepackage{graphicx}
\usepackage[breaklinks, colorlinks=true]{hyperref}
\usepackage{bm}

\usepackage{physics}

\newcommand{\Nc}{N_{\text{c}}}
\newcommand{\iomega}{\Omega_{\rm I}}
\newcommand{\tiomega}{\tilde{\Omega}_{\rm I}}

\newcommand{\bk}{{\vec{k}}}
\newcommand{\Tc}{T_{\text{c}}}
\newcommand{\bphi}{{\bm\phi}}
\newcommand{\balpha}{{\bm\alpha}}

\newcommand{\Z}{\mathbb{Z}}

\begin{document}
\title{Perturbative confinement in
thermal Yang-Mills theories induced by imaginary angular velocity}

\author{Shi Chen}
\email{s.chern@nt.phys.s.u-tokyo.ac.jp}
\affiliation{Department of Physics, The University of Tokyo, 
  7-3-1 Hongo, Bunkyo-ku, Tokyo 113-0033, Japan}

\author{Kenji Fukushima}
\email{fuku@nt.phys.s.u-tokyo.ac.jp}
\affiliation{Department of Physics, The University of Tokyo, 
  7-3-1 Hongo, Bunkyo-ku, Tokyo 113-0033, Japan}

\author{Yusuke Shimada}
\email{yshimada@nt.phys.s.u-tokyo.ac.jp}
\affiliation{Department of Physics, The University of Tokyo, 
  7-3-1 Hongo, Bunkyo-ku, Tokyo 113-0033, Japan}

\begin{abstract}
  We perturbatively compute the Polyakov loop potential at high temperature with finite imaginary angular velocity.
  This imaginary rotation does not violate the causality and the thermodynamic limit is well defined.
  We analytically show that the imaginary angular velocity induces the perturbatively confined phase and serves as a new probe to confinement physics.
  We discuss a possible phase diagram that exhibits adiabatic continuity from the perturbative confinement to the confined phase at low temperature.
  We also mention subtlety in the analytical continuation from imaginary to real angular velocity by imposing a causality bound.
\end{abstract}
\maketitle

\paragraph{Introduction:}
Confinement of quarks and gluons in quantum chromodynamics (QCD) has been a long-standing problem.
There are traditional strategies to idealize the problem.
One can take special limits such as the strong-coupling limit~\cite{Wilson:1974sk,Polyakov:1978vu}, the large-$\Nc$ limit~\cite{tHooft:1973alw}, etc. to scrutinize the confinement mechanism in a nonperturbative and yet analytical way.
Such deformations of the theory belong to a category which we would call the \textit{QCD-like theory approach}.
Examples of QCD-like theories include
holographic QCD models~\cite{Babington:2003vm,Kruczenski:2003uq,Sakai:2004cn} and
a supersymmetric extension~\cite{Seiberg:1994rs,Seiberg:1994aj,Seiberg:1994pq,Davies:1999uw}.
In particular, an interesting idea of the ``adiabatic continuity'' on a small circle~\cite{Davies:1999uw,Poppitz:2012sw,Aitken:2017ayq} based on the Polyakov mechanism~\cite{Polyakov:1976fu} has been recognized.

Another stream of research toward the confinement mechanism is the introduction of external parameters corresponding to extreme environments such as the temperature $T$, the density or the chemical potential $\mu$, the magnetic field $B$, and so on, which we call the
\textit{extreme QCD approach}~\cite{Rajagopal:2000wf,Fukushima:2011jc}.
QCD at 
high temperature is perturbatively tractable and the loop calculation of the confinement order parameter, i.e., the Polyakov loop, has led to
{a} deconfined phase~\cite{Weiss:1980rj,Weiss:1981ev,Gross:1980br,KorthalsAltes:1993ca,Gocksch:1993iy} (for a review, see Ref.~\cite{Fukushima:2017csk}).
Generally speaking, extreme environments provide an energy scale greater than the QCD scale, so that the perturbative calculation in favor of 
{deconfinement} 
is justified.
The perturbative analysis breaks down with decreasing $T$/$\mu$/$B$, and it is usually impossible to go into the confinement regime.
Nevertheless, one may perceive a precursory tendency of
{a confinement phase}
transition
(see also Refs.~\cite{Vuorinen:2006nz,Fukushima:2013xsa} for approaches to enforce confinement to come closer to the transition point).
One could also employ other external probes like the electric field $E$~\cite{Yamamoto:2012bd}, the isospin chemical potential $\mu_{\text{iso}}$~\cite{Son:2000by,Son:2000xc}, the scalar curvature $R$~\cite{Inagaki:1997kz,Flachi:2014jra}, the rotational angular velocity $\omega$~\cite{Yamamoto:2013zwa,Chen:2015hfc,Jiang:2016wvv,Chernodub:2020qah,Chen:2020ath,Braguta:2020biu,Braguta:2021jgn,Fujimoto:2021xix}, and their mixtures~\cite{Flachi:2017vlp,Zhang:2018ome}.

Since an extraordinary value of $\omega\sim 10^{22}\,\mathrm{s}^{-1}$ was reported in the heavy-ion collision experiment~\cite{STAR:2017ckg}, 
the effect of as large $\omega$ as the QCD scale has been attracting theoretical and experimental interests.
Model calculations implied similarity between the angular velocity and the chemical potential~\cite{Chen:2015hfc}, which was summarized in a form of the QCD phase diagram on an $\omega$-$T$ plane~\cite{Jiang:2016wvv}.
The lattice-QCD simulation suffers from the sign problem at finite $\omega$ in the same way as the finite-$\mu$ case.
However,
the lattice-QCD simulation with the analytical continuation from the imaginary angular velocity $\iomega$ to $\omega$ via $\iomega=-i\omega$ is feasible because $\iomega$ does not cause the sign problem~\cite{Yamamoto:2013zwa,Braguta:2020biu,Braguta:2021jgn}.
Here, we shall emphasize that such a system with imaginary rotation $\iomega$ is quite intriguing on its own.
With explicit calculations we show that the pure Yang-Mills (YM) theory at sufficiently large $\iomega$ goes through a 
{confinement}
phase transition 
even perturbatively.

One might also obtain the perturbatively confining phase by adding finely tuned quark contents, such as a massless adjoint Dirac fermion with imaginary chemical potential, $\mu_{\mathrm{I}}=\pi$.
However, the present work is the very first report of perturbative confinement from purely gluonic loops, to the best of our knowledge.
In the literature, confining mechanisms that are essentially gluonic are all nonperturbative.
They hinge on either 
semiclassical contribution~\cite{Polyakov:1976fu,Davies:1999uw,Poppitz:2012sw,Aitken:2017ayq}, 
lattice regularization~\cite{Wilson:1974sk,Polyakov:1978vu}, electricity-magnetism duality~\cite{Seiberg:1994rs,Seiberg:1994aj,Seiberg:1994pq}, 
or dressed ghost/gluon propagators~\cite{Gribov:1977wm,Kugo:1979gm,Zwanziger:1993dh,Braun:2007bx}.
In contrast to these preceding works, surprisingly, we find that purely gluonic confinement in (3+1)D is possible without invoking any nonperturbative machinery.
\vspace{0.5em}

\paragraph{Polyakov loop potential with imaginary rotation:}
In the pure YM theory, we perform the one-loop calculation to find the Polyakov loop potential which is often called the
Gross-Pisarski-Yaffe-Weiss (GPY-W)
potential~\cite{Gross:1980br,Weiss:1980rj,Weiss:1981ev}. 
{Under imaginary rotation by $\iomega$~\cite{Chernodub:2020qah}, we find a system of Euclidean cylindrical coordinates, i.e., $x^\mu=(\tau,\theta, r ,z)$, with the flat metric $g_{\mu\nu}=\mathrm{diag}\{1,r^2,1,1\}$ and the following boundary condition:}
\begin{equation}\label{eq:BC}
    (\tau,\,\theta,\,r,\,z)\sim(\tau+\beta,\,\theta-\tiomega,\,r,\,z)\,,
\end{equation}
where $\beta=1/T$ is inverse temperature and $\tiomega:= \iomega/T$.
Clearly, $\tiomega$ and $\tiomega+2\pi$ describe the same geometry.
In the presence of the Polyakov loop background,
$\partial_\tau$ is replaced by the covariant derivative $D_\tau$ as
\begin{equation}
    D_\tau = \partial_\tau + i
    \frac{\bphi\cdot\bm H}{\beta}\,.
    \label{eq:D_tau}
\end{equation}
The $\mathfrak{g}$-valued vector $\bm H$ is an orthonormal basis of a Cartan subalgebra of $\mathfrak{g}$, the Lie algebra of the gauge group. Thus the Polyakov loop is labeled with a real vector $\bphi$.
We take homogeneous $\bphi$ backgrounds because they are the classical vacua even in the presence of $\iomega$. 

To perform the one-loop integral, we need to diagonalize the fluctuation operator. For ghosts, it is the scalar Laplacian,
$- D^2_{\mathrm{s}} = - D_\tau^2 - r^{-1}\partial_ r ( r \partial_ r ) -  r^{-2}\partial_\theta^2 - \partial_z^2$.
We solve the eigenequation, $-D^2_{\mathrm{s}}\Phi=\lambda\Phi$, with the twisted boundary condition~\eqref{eq:BC} to find the spectrum.
Since we are merely interested in a potential of $\bphi$, we drop the eigenmodes that commute with $\bm H$. Then we find,
\begin{equation}
    \Phi_{n,m,\bk,\balpha}(x) = \frac{E_{\balpha}}{\sqrt{2\pi\beta}}\,
    e^{i\left[\left(\frac{2\pi n}{\beta} + \iomega m \right)\tau + m\theta + k_z z\right]}J_{m}(k_\perp r )\,.
    \label{eq:scalarMode}
\end{equation}
Here, $n,m\in\Z$,
$\bk:=(k_\perp,k_z)\in\mathbb{R}^+\!\times\!\mathbb{R}$
and
$\balpha$'s are roots of $\mathfrak{g}$.
The eigenvalues are given by
\begin{equation}
    \lambda_{n,m,\bk,\balpha}=\left(\frac{2\pi n + \bphi\cdot\balpha}{\beta} + \iomega m \right)^2 + |\bk|^2\,.
    \label{eq:E}
\end{equation}
We can generalize the above calculation to the covariant vector fields, for which the Laplacian is a $4\times 4$ matrix given by 
\begin{equation}
    - D^2_{\mathrm{v}} = 
    \begin{pmatrix}
    - D^2_{\mathrm{s}} & 0  & 0 & 0 \\
    0 & - r D^2_{\mathrm{s}} r^{-1} + r^{-2} & -2 r^{-1}\partial_\theta & 0 \\
    0 & 2 r^{-3}\partial_\theta & - D^2_{\mathrm{s}} + r^{-2} & 0 \\
    0 & 0 & 0 & - D^2_{\mathrm{s}}
    \end{pmatrix}\,.
\end{equation}
Its eigenvalues are the same as Eq.~\eqref{eq:E} but its eigenmodes come with a degeneracy of four polarizations.
The unphysical (non-transverse) polarizations are simply replicas of the scalar mode~\eqref{eq:scalarMode}, i.e.,
$\Xi_{n,m,\bk,\balpha}^{(i)}(x) = \Phi_{n,m,\bk,\balpha}(x)\,\xi^{(i)}$,
where $\xi^{(1)}:=(1,0,0,0)^T$ 
and
$\xi^{(2)}:=(0,0,0,1)^T$.
The loops of these unphysical eigenmodes are canceled by the ghost loop.
The physical transverse eigenmodes have nontrivial tensorial structure with $m$ shifted by the helicity of the vector fields as
\begin{equation}\label{eq:vectorMode}
\begin{split}
    \Xi_{n,m,\bk,\balpha}^{(\pm)}(x)=&\\
     \frac{E_{\balpha}\,\xi^{(\pm)}}{2\sqrt{\pi\beta}} &e^{i\left[\left(\frac{2\pi n}{\beta} + \iomega m\right)\tau + m\theta + k_z z\right]}J_{m\pm1}(k_\perp r) \,,
\end{split}
\end{equation}
where $\xi^{(\pm)}:=(0, r, \pm i, 0)^T$.

After performing the Matsubara summation and dropping the ultraviolet divergence independent of $\bphi$, 
we find the following expression for the effective potential:
\begin{align}
    & V = \frac{T}{4\pi^2} \sum_{\balpha}\sum_{m\in\Z}
    \int_0^\infty \! k_\perp dk_\perp \int_{-\infty}^{\infty} \! dk_z 
    \Bigl[J^2_{m-1}(k_\perp r)
    \notag\\
    &\qquad +J^2_{m+1}(k_\perp r)\Bigr]
    \mathrm{Re} \ln\qty[1\!-\!e^{-\qty(|\bk|-i\iomega m)/T +i\bphi\cdot\balpha}].
    \label{eq:Vinteg}
\end{align}
Interestingly, we can analytically perform the summation and integrals using the power series:
$\displaystyle \ln(1-z)=-\sum_{l=1}^\infty \frac{z^l}{l}$ which converges for $|z|\leq1$, $z\neq1$. 
We then obtain a simple expression,
\begin{equation}\label{eq:Vfull}
    V(\bphi;\tiomega) = -\frac{2T^4}{\pi^2}\sum_{\balpha}
    \sum_{l=1}^\infty
    \frac{\cos(l\bphi\cdot\balpha) \cos(l\tiomega)}
    {\Bigl\{ l^2 + 2\tilde{r}^2 \bigl[1-\cos (l\tiomega)\bigr]\Bigr\}^2}\,,
\end{equation}
where we introduced dimensionless $\tilde{r}:=r T$.
{
At $\tiomega = 0 \mathrm{~mod~} 2\pi$, Eq.~\eqref{eq:Vfull} loses its $r$-dependence and recovers the well-known GPY-W potential~\cite{Gross:1980br,Weiss:1980rj,Weiss:1981ev}.

For a concrete reference, we shall focus on the rotation center, $\tilde{r}\sim0$, in this Letter.  
However, we note that, at $\tiomega = \pi \mathrm{~mod~} 2\pi$,
the $r$-dependent potential in Eq.~\eqref{eq:Vfull} results in homogeneous $\bphi$-vacua for SU(2).
We shall shortly reveal that the most nontrivial physics exactly inhabits this homogeneous region, so that we can extend our conclusion to $\tilde{r}\neq 0$.
}
At $\tilde{r}=0$ we can complete the $l$ summation to find:
\begin{equation}
    V(\bphi;\tiomega)|_{\tilde{r}= 0} =
    \frac{\pi^2 T^4}{3} \sum_{\balpha}\!\sum_{s=\pm 1}\!
    B_4\biggl( \Bigl(\frac{\bphi\cdot\balpha+s\tiomega}{2\pi}\Bigr)_{\text{mod 1}}\biggr).
    \label{eq:V_r=0}
\end{equation}
Here $B_4(x)=x^4-2x^3+x^2-\frac{1}{30}$ is the 4th Bernoulli polynomial. 
Equation~\eqref{eq:V_r=0} has quite rich physical contents despite its simple appearance.
\vspace{0.5em}

\paragraph{Perturbative confinement phase transition:}
We now investigate the evolution of the Polyakov loop potential with increasing $\tiomega$.
Let us start with the simplest SU(2) gauge group.
Here we define $\phi:=\bphi\cdot\balpha$ for the only positive root $\balpha$.
Modulo periodicities and the Weyl group, $\phi$ runs in $[0,2\pi]$ and the $\Z_2$ center symmetry acts as $\phi\to2\pi-\phi$.

\begin{figure}
    \centering
    \includegraphics[width=0.85\columnwidth]{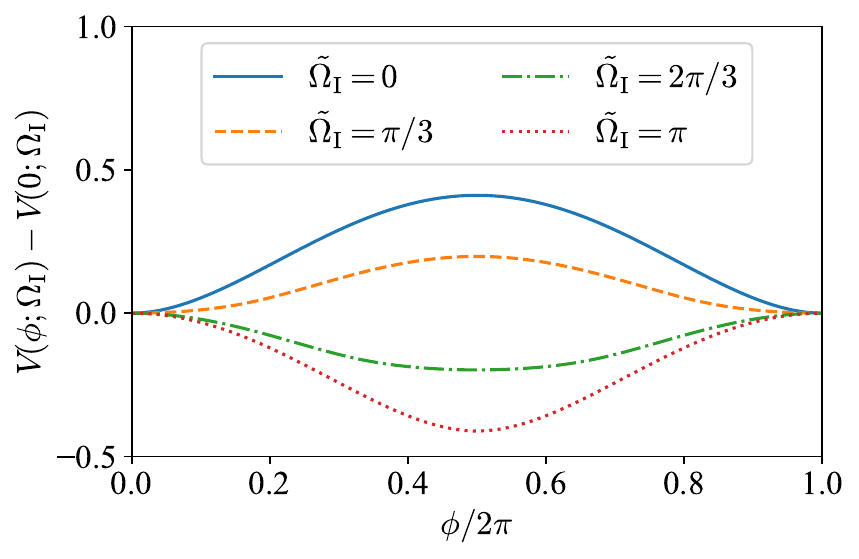}
    \caption{Evolution of the Polyakov loop potential (made dimensionless with $T^4$) for $\tiomega=0, \pi/3, 2\pi/3, \pi$ in the color SU(2) case at $\tilde{r}=0$.}
    \label{fig:potential}
\end{figure}

\begin{figure}
    \centering
    \includegraphics[width=0.85\columnwidth]{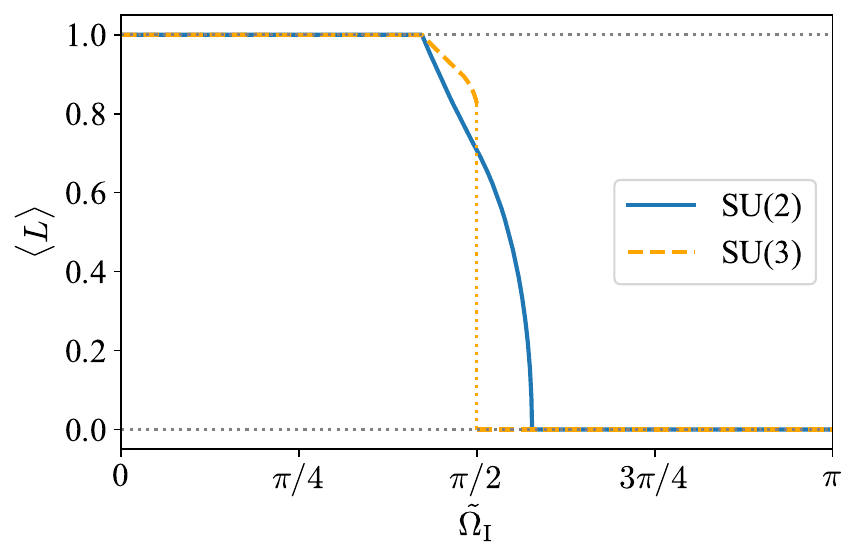}
    \caption{The expectation value of the fundamental Polyakov loop, normalized by the representation dimension, as a function of $\tiomega$ for SU(2) (solid line) and SU(3) (dashed line) at $\tilde{r}=0$.}
    \label{fig:polyakov}
\end{figure}

Figure~\ref{fig:potential} shows the evolution of the Polyakov loop potential in terms of $\phi/2\pi$ with increasing $\tiomega$ at $\tilde{r}=0$.
The solid curve in Fig.~\ref{fig:potential}
for $\tiomega=0$ reproduces the center breaking GPY-W potential with minima located at $\phi=0$ and $2\pi$.
The positive curvature around the minima then corresponds to the Debye screening mass that stabilizes the deconfined phase at high temperature~\cite{Rebhan:1994mx}.
We clearly see that the curvature is suppressed as $\tiomega$ gets larger, and eventually the sign of the curvature flips around $\tiomega\simeq \pi/2$.
Then, the potential minima deviate from the deconfined vacua and the confined vacuum at $\phi=\pi$ is energetically favored.
We can visualize this phase transition by plotting $\langle L\rangle$, 
the expectation value of the fundamental Polyakov loop $L$,
as a function of $\tiomega$ as shown in Fig.~\ref{fig:polyakov}.
We see that $\langle L\rangle$ starts to decrease from $\tiomega=(1-1/\sqrt{3})\pi$.
The dropping curve hits $\langle L\rangle=0$ at $\tiomega=\pi/\sqrt{3}$,
indicating a second-order confinement phase transition.

We can intuitively understand the confining force at $\tiomega=\pi$ from the twisted geometry~\eqref{eq:BC}.
It assigns the antiperiodic boundary condition to all odd-$m$ transverse modes~\eqref{eq:vectorMode}.
But these modes still obey bosonic statistics such that their loops have no overall sign of $-1$.
At $r=0$, only the modes of $m=\pm1$ contribute.
Such antiperiodic gluons reverse the one-loop potential just in analogy to periodic gluinos.

\begin{figure}
    \centering
    \includegraphics[width=0.85\columnwidth]{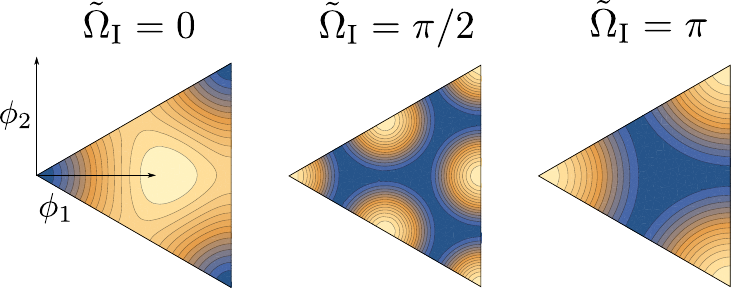}
    \caption{Polyakov loop potential for the SU(3) case.  The light (dark) color indicates the region of larger (smaller) potential values.}
    \label{fig:su3}
\end{figure}

We move on to the SU(3) case.
The positive roots are $\balpha_1=(1,0)$, $\balpha_2=(1/2, \sqrt{3}/2)$,
and $\balpha_3=(1/2, -\sqrt{3}/2)$.
Accordingly, the order parameter has two components, namely,
$\bphi=(\phi_1,\phi_2)$. Modulo periodicities and the Weyl group, $\bphi$ runs in a triangular region spanned by the vertices $(0,0)$, $(2\pi, 2\pi/\sqrt{3})$, and $(2\pi,-2\pi/\sqrt{3})$, as drawn in Fig.~\ref{fig:su3}. 
The points in this triangle bijectively represent conjugacy classes of SU(3).
The $\Z_3$ center symmetry acts on this equilateral triangle as its rotational geometry symmetry.

We show the SU(3) potential height in the form of the contour plot in Fig.~\ref{fig:su3}.
The lighter (darker) color indicates the region of larger (smaller) potential values.
The left in Fig.~\ref{fig:su3} presents the potential profile at $\tiomega=0$.
The minima are located at $(0,0)$ and its center symmetry images, which signifies the spontaneous breaking of center symmetry.
With increasing $\tiomega$, these minima depart from the conventional vacua as we observed in the SU(2) case.
A crucial difference of SU(3) from the SU(2) case is, as shown in the middle of Fig.~\ref{fig:su3}, the center symmetric point $(4\pi/3,0)$ is pushed down and eventually at $\tiomega=\pi/2$ we see degeneracy between three shifted deconfined vacua and the center symmetric point.
The degeneracy indicates a first-order phase transition, and
the center symmetric (confining) state is energetically favored for $\tiomega=\pi$ as shown in the right of Fig.~\ref{fig:su3}.
We can also visualize this first-order nature by plotting $\langle L\rangle$ as shown in Fig.~\ref{fig:polyakov}.
Clearly, we see a sudden jump of $\langle L\rangle$ at $\tiomega=\pi/2$.
This difference in the order of the phase transition between SU(3) and SU(2) is consistent with the universality class argument~\cite{Svetitsky:1982gs}.

Our formulae hold for any semisimple Lie algebra. 
We can show that, for any simply-connected compact gauge group with a nontrivial center, Eq.~\eqref{eq:V_r=0} at $\tiomega=\pi$ always favors a center symmetric vacuum.
For example,
Spin(5) also exhibits a first-order confinement phase transition at $\tiomega=\pi/2$.
A more interesting case is $G_2$ which has no center symmetry.
Consistently, we observed no phase transition; the location of its potential minimum just moves continuously as a function of $\tiomega$.
\vspace{0.5em}

\paragraph{Phase diagram and adiabatic continuity:}
It is an intriguing question whether, on the $\tiomega$-$T$ plane, the perturbatively confined phase we found above is smoothly connected to the conventional confined phase.
Although our loop calculations cannot constrain the low-$T$ physics, the Kugo-Ojima-Gribov-Zwanziger (KOGZ) mechanism~\cite{Gribov:1977wm,Kugo:1979gm,Zwanziger:1993dh} still allows us to grasp some hints
{as follows.

The ghost contribution to the one-loop potential that favors confinement is just negative of Eq.~\eqref{eq:Vinteg} with $J_{m\pm1}$ replaced by $J_m$. 
Around $r=0$, because only the $m=0$ component remains, this ghost contribution does not depend on $\tiomega$.
At high $T$, perturbatively, this ghost potential cancels out with the contribution from unphysically polarized gluons.
At low $T$, the KOGZ mechanism asserts that the ghost propagator is nonperturbatively enhanced at infrared (nearly divergently)~\cite{Braun:2007bx}.
Therefore, as the system is cooled down, the ghost makes the system increasingly confining, uniformly for any $\tiomega$.
}

\begin{figure}
    \centering
    \includegraphics[width=0.75\columnwidth]{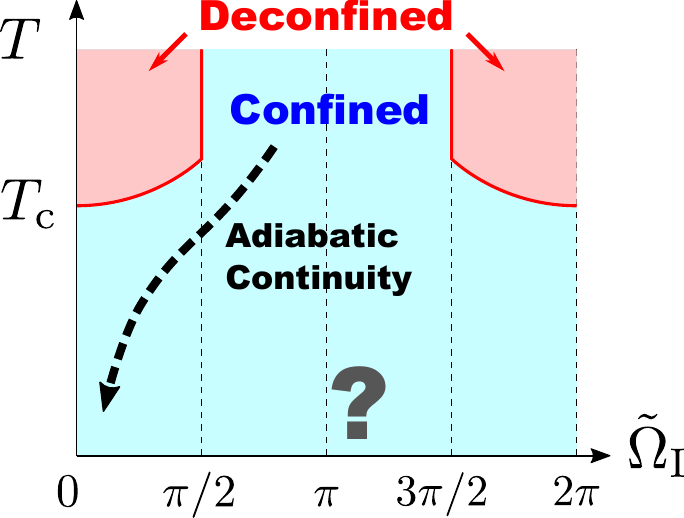}
    \caption{Conjectured phase diagram on the $\tiomega$-$T$ plane around the rotation axis, $\tilde{r}=0$, for the SU(3) case. Solid curves represent the phase transition.  
    }
    \label{fig:conjecture}
\end{figure}

Based on the arguments above, taking the SU(3) case, we sketch a phase diagram as shown in Fig.~\ref{fig:conjecture}. 
A remarkable feature is the uniform confinement for all $T$ around $\tiomega=\pi$.
Let us elaborate our speculated physics in this region.
Like the Debye mass in the deconfined phase, the string tension $\sigma$ in the confined phase is indicated by the curvature of the effective potential around minima.
Thus we have $\sigma\sim g^2 T^2$ at very high $T$.
As we cool down the system, this perturbative string tension decreases until it reaches $\sigma\sim\Lambda_{\text{YM}}^2$,
where $\Lambda_{\text{YM}}$ is the dynamical scale from conformal anomaly. 
At smaller temperature the system goes into the nonperturbative confining region and the string tension is kept about $\sigma\sim\Lambda_{\text{YM}}^2$. 

In our speculated phase diagram in Fig.~\ref{fig:conjecture}, our perturbatively confined phase around $\tiomega=\pi$ is connected to the conventional confined phase at $\tiomega=0$ without a phase transition.
Then, by \textit{adiabatic continuity}, we can study quite a few features of the conventional confinement phase transition even using loop calculations.
For example, we already predicted the scaling of the string tension, the order of the phase transition, etc.
It 
{would be}
fascinating to examine our conjectured phase diagram, as well as the realization of adiabatic continuity, by feasible nonperturbative methods such as the lattice numerical simulation.

\vspace{0.5em}

\paragraph{Analytical continuation to real rotation:}
We finally apply our results to real rotation.
It is customary in the literature to study the real rotation effect by the analytical continuation from $\iomega$ to $\omega$~\cite{Braguta:2020biu,Braguta:2021jgn}.
For example, once $\Tc(\tiomega)$ is known, then $\Tc(\omega)$ is inferred from the replacement of $\iomega^2=-\omega^2$.
However, we explicate that such a procedure might be problematic using our perturbative expression.

For any {complex $\tiomega$ outside the real axis}, Eq.~\eqref{eq:Vinteg} yields singularity at some $\bphi$.
In fact, our derivation of Eq.~\eqref{eq:Vfull} is valid for real $\tiomega$ only.  A nonzero $\mathop{\mathrm{Im}}\iomega$ would drive the Maclaurin series of $\ln(1-z)$ out of its convergence radius. 
If we na\"{i}vely perform the analytical continuation to Eq.~\eqref{eq:Vfull}, we would also encounter the following problem.
For $\tilde{r} > 0$ Eq.~\eqref{eq:Vfull} is analytical everywhere except on the imaginary $\tiomega$ axis.
There, infinitely many poles are accumulated around $\tiomega=0$.
As for $\tilde{r}=0$, the poles are gone, but the infinite summation just blows up for nonreal $\tiomega$.

\begin{figure}
    \centering
    \includegraphics[width=.9\linewidth]{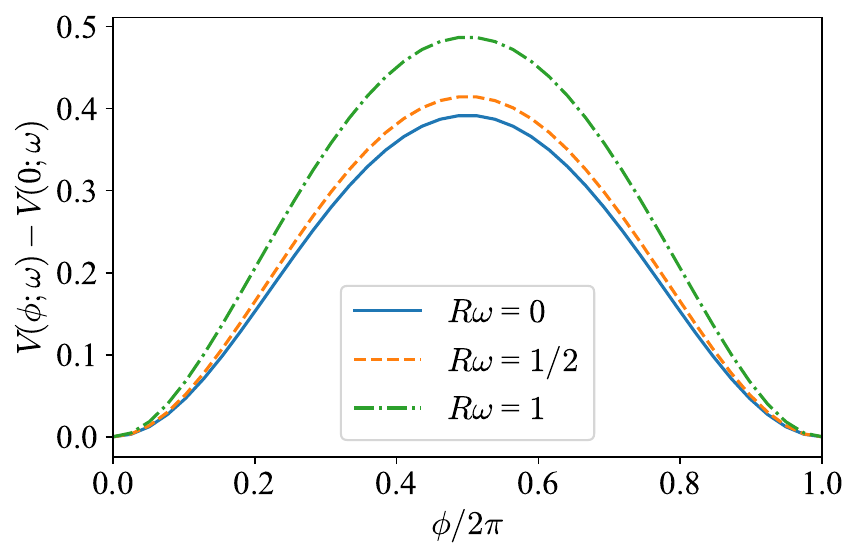}
    \caption{Evolution of the Polyakov loop potential (made dimensionless with $T^4$) for $\omega R = 0, 1/2, 1$ in the color SU(2) case at $\tilde{r}=0$.  Our choice of parameters is: $R = 10\,\text{GeV}^{-1}$ and $T=0.15\,\text{GeV}$.}
    \label{fig:potential_real}
\end{figure}

The physical origin of these singularities is clear.
At finite angular velocity $\omega$, the longwave modes with $k_\perp \lesssim\omega$ violate the causality so we should introduce an infrared cutoff, i.e.,
$r\omega$ must not exceed the unity.
Let us set the system size as $r \le R$ with $R\omega\le 1$.
This discretizes the momentum $k_\perp$ such that $k_\perp R$ is a zero of the Bessel functions. 
Here, we denote the $\kappa$-th zero of $J_\nu(\xi>0)$ as $\xi_{\nu,\kappa}$.
The phase space integration in Eq.~\eqref{eq:Vinteg} is replaced as follows:
\begin{equation}
  \begin{split}
  &\int_0^\infty k_\perp dk_\perp \ J_m^2(k_\perp r) \, f(k_\perp) \\
  &\to \sum_{\kappa = 1}^\infty \frac{2}{R^2 J_{m+1}^2(\xi_{m,\kappa})} J_m^2\qty(\frac{\xi_{m,\kappa}\, r}{R})\, f\qty(\frac{\xi_{m,\kappa}}{R})\,.
  \end{split}
  \label{eq:kmodify}
\end{equation}
We have performed the numerical integration and summation of Eq.~\eqref{eq:Vinteg} for real $\omega$ with Eq.~\eqref{eq:kmodify} substituted.
We cut off the sum over $m$, $\kappa$ and the $k_z$ integration by sufficiently large numbers and confirm the convergence.

Figure~\ref{fig:potential_real} shows the evolution of the SU(2) Polyakov loop potential with increasing $\omega$ at $\tilde{r} = 0$.
We chose the parameters as $R = 10\,\text{GeV}^{-1} \;(\simeq 2\,\text{fm})$
and $T=0.15\,\text{GeV}$.
The potential minima are located at $\phi=0 \mathrm{~mod~} 2\pi$ for any $\omega$, so that the system stays in the deconfined phase.
Yet, we can quantify the effect of $\omega$ onto the vacuum stability by the potential curvature around the minimum which represents the Debye screening mass squared.
As we see in Fig.~\ref{fig:potential_real},
the curvature increases with increasing $\omega$, and this means that rotation favors deconfinement.
This behavior makes a contrast to the results from Refs.~\cite{Braguta:2020biu,Braguta:2021jgn}, while the recent lattice results from Ref.~\cite{Chernodub:2022veq} support our conclusion.
We found that the Polyakov loop potential even around $r=0$ is sensitive to the system size $R$ and the boundary treatments.
We are now investigating the effects of boundary and axial symmetry breaking on the lattice to clarify the validity range of analytical continuation to real rotation.
\vspace{0.5em}

\paragraph{Outlook:}
An intriguing and immediate extension of our work would be the lattice simulation to explore the whole $\tiomega$-$T$ phase structure as conjectured in Fig.~\ref{fig:conjecture}, {complementary to preceding efforts~\cite{Braguta:2020biu,Braguta:2021jgn} with a boundary condition}.
Actually, we can have the lattice simulation at our fingertips for $\tiomega=\pi/2$ and $\pi$ 
by moving to the Cartesian coordinates, $(\tau,x,y,z)$, 
where the boundary condition~\eqref{eq:BC} reduces to 
$(\tau,x,y,z)\sim(\tau+\beta,y,-x,z)$ at $\tiomega=\pi/2$ and 
$(\tau,x,y,z)\sim(\tau+\beta,-x,-y,z)$ at $\tiomega=\pi$, respectively.
Therefore, for $\tiomega=\pi/2$ and $\pi$, we do not have to deal with nontrivial geometry, but just take the square lattice and the Cartesian spacetime only with a twisted thermal boundary condition.
Once we manage to know the physics at $\tiomega=\pi/2$ and $\pi$ for various temperatures, we can justify our speculated scenario of adiabatic continuity in Fig.~\ref{fig:conjecture}.
{Another exciting extension is to include the fundamental/adjoint quark contributions and to discuss a relation to chiral symmetry.}
\vspace{0.5em}


\begin{acknowledgments}
The authors thank
Maxim~Chernodub,
Yuki~Fujimoto,
and
Arata~Yamamoto
for useful discussions.
They also thank
Yuya~Tanizaki
for pointing out that the deconfined phase at low $T$ is unlikely, which strengthens our speculation.
This work was supported by Japan Society for the Promotion of Science
(JSPS) KAKENHI Grant No.\ 21J20877 (S.C.),
19K21874 (K.F.),
22H01216 (K.F.).
\end{acknowledgments}        

\bibliography{rotatingweiss}
\bibliographystyle{apsrev4-2}
\end{document}